# Effect of Dzyaloshinskii–Moriya interaction on magnetic vortex

Y. M. Luo[1], C. Zhou[1], C. Won[2] and Y. Z. Wu[1,*]

[1]*State Key Laboratory of Surface Physics, Department of Physics, and Advanced Materials Laboratory, Fudan University, Shanghai 200433, People's Republic of China*

[2]*Department of Physics, Kyung Hee University, Seoul 130-701, Korea*

Abstract

The effect of the Dzyaloshinskii–Moriya (DM) interaction on the vortex in magnetic microdisk was investigated by micro-magnetic simulation based on the Landau–Lifshitz–Gilbert equation. Our results show that the DM interaction modifies the size of the vortex core, and also induces an out-of-plane magnetization component at the edge and inside the disk. The DM interaction can destabilizes one vortex handedness, generate a bias field to the vortex core and couple the vortex polarity and chirality. This DM-interaction-induced coupling can therefore provide a new way to control vortex polarity and chirality.

1. Introduction

A novel antisymmetric exchange coupling [1,2] called the Dzyaloshinskii–Moriya (DM) interaction has recently attracted great interest. The DM interaction arises from spin-orbit scattering of electrons in an inversion asymmetric crystal field, and it exists in systems with broken inversion symmetry, such as in specific metallic alloys with B20 structure [3,4,5,6,7] and at the surface or interface of magnetic multi-layers [8,9,10]. The existence of the DM interaction can induce chiral spin structures such as skrymion [3-10], unconventional transport phenomena [11,12,13], and exotic dynamic properties [14,15,16], many of which stimulated interest in fundamental magnetism studies and provided new possibilities for the development of future spintronic devices.

Besides the DM-interaction-induced effects in bulk materials and thin films, the practical consequences of the DM interaction in confined structures such as magnetic nanodisks and nanostripes have begun to attract increasing interest [17,18,19,20]. The stable magnetic configuration in sub-micrometer scale magnetic microdisk is a magnetic vortex, which can be characterized by an in-plane curling magnetization (chirality) and a nanometer-sized central region with an out-of-plane magnetization (polarity) [21,22]. The chirality can be clockwise ($C=-1$) or counterclockwise ($C=1$), and the polarity can be up ($P=1$) or down ($P=-1$). The vortices can also be classified as left-handed vortex ($CP=-1$) and right-handed vortex ($CP=1$) [23,24]. In the classic model, the vortex chirality and polarization are not coupled and can be switched independently, but one recent experiment indicated that the DM interaction can break this symmetry [24]. A few theoretical investigations were carried out to explore the influence of the DM interaction on magnetic vortices [19,20]. Through Monte-Carlo simulation, Kwon et. al showed that even without the dipolar interaction the DM interaction could induce the vortex structure in a nanodisk with the radium less than 30nm [20]. Butenko et. al found that DM coupling can considerably change the size of vortices [19], but ignored its effect on the magnetization at the disk edge which exists in the real magnetic vortex system. Usually the experimental studies on the magnetic vortex were performed in the magnetic disk with the diameter around the micrometer size [21-24].

In order to have a measurable study on the effect of DM interaction on the magnetic vortex before the further experimental study, we performed a micro-magnetic simulation together with the existing demagnetization field on a magnetic microdisk. Our results showed that the existence of the DM interaction not only shrinks or broadens the vortex core, but it also induces an out-of-plane magnetization component both at the edge and at the disk plane, which has a linear dependence on the DM interaction strength. Moreover, we found that the DM interaction can induce a bias field on the vortex core, so that a clear bias effect can be observed through the vortex core switching process. Thus the DM interaction can couple the vortex chirality and polarity, which provides a new possibility for manipulating the vortex chirality and polarity together.

2. Equations and Methods

In our simulation, the spin system is described by a 2-dimentional (2D) square lattice, with a local magnetic moment $|\vec{m}| = M_S$ at each site. This model includes the ferromagnetic exchange interaction, the DM interaction, the magnetic dipole interaction, and Zeeman coupling, and the magnetocrystalline anisotropy was ignored. The Hamiltonian can be written as:

$$E = -J\sum \vec{m}_i \cdot \vec{m}_j + D\sum \vec{r}_{ij} \cdot (\vec{m}_i \times \vec{m}_j) - \mu_0 \vec{H} \cdot \sum \vec{m}_i - \frac{1}{2}\mu_0 \sum \vec{m}_i \cdot \vec{H}_d \tag{1}$$

where $J$, $D$, $\vec{r}_{ij}$, $\mu_0$, $\vec{H}$ and $\vec{H}_d$ denote the exchange constant, the DM constant, the distance vector between the spin sites $i$ and $j$, magnetic permeability, the external magnetic field and demagnetization field, respectively. $\vec{H}_d$ is computed from the magnetization distribution through the magnetostatic equations [25, 26]. In this study, we considered the DM interaction as those in the materials with B20 structure [3-7] that can induce the helical spiral stripes. The spiral period is determined by the ratio $J/D$ [6], and the helical direction is determined by the sign of $D$: negative $D$ produces a left-handed helical structure, and positive $D$ produces a right-handed helical structure [9].

Generally, the spin configuration can be simulated by numerically solving the Landau–Lifshitz–Gilbert (LLG) equation:

$$\frac{d\vec{m}_i}{dt} = -|\gamma|\vec{m}_i \times \vec{H}_{eff} - \frac{\alpha}{M_s}\vec{m}_i \times \frac{d\vec{m}_i}{dt} \qquad (2)$$

with the total effective field $\vec{H}_{eff}$, the Gilbert gyromagnetic ratio $\gamma$ and the damping constant $\alpha$. The total effective field includes the exchange field, the dipole field, the DM field, and the external field, and can be written as:

$$\vec{H}_{eff} = -\frac{1}{\mu_0}\frac{\partial E}{\partial \vec{m}_i} \qquad (3)$$

where $E$ is the total energy of the system as expressed in equation (1). Typically, the effective field induced by the DM interaction can be expressed as:

$$\vec{H}_{DM} = -\frac{1}{\mu_0}\frac{\partial D\sum \vec{r}_{ij}\cdot(\vec{m}_i \times \vec{m}_j)}{\partial \vec{m}_i} = -\frac{D}{\mu_0 M_s}\sum(\vec{m}_j \times \vec{r}_{ij}) \qquad (4)$$

In the continuous limit, the DM interaction can be written as [27]

$E_{DM} = -D\vec{m}\cdot(\nabla \times \vec{m})$, thus its effective field can be expressed as:

$$\vec{H}_{DM} = -\frac{2D}{\mu_0 M_s}(\nabla \times \vec{m}) \qquad (5)$$

We realized the simulation by adding a DM interaction module into the standard micro-magnetic simulation software OOMMF [28]. The micro-magnetic simulation is usually considered to directly compare with the real material system, so in the simulation we chose the typical parameters [29], such as $M_s = 8\times 10^5 A/m$, the exchange stiffness $A = 1.3\times 10^{-11} J/m$ and $\alpha = 0.01$, but the DM interaction was regarded as a tuning parameter. In the simulation, the nanodisk diameter is $502 nm$, and the thickness is $50 nm$; in this case the stable magnetic configuration is vortex.

We only report the simulation results with the unit cell of $2\times 2\times 50\, nm^3$, thus the center spin can point to the direction exactly perpendicular to the film plane. We also did the 2D calculation with a unit cell size of $1\times 1\times 50\, nm^3$ on a diameter of 501nm, and obtained the same results. So the chosen unit cell of $2\times 2\times 50\, nm^3$ is accurate enough for the current study. In order to make sure that the DM interaction will not significantly change the Neumann boundary condition used in the OOMMF code [30], we tested the calculation on 2×2 disk arrays with 100nm separation, and obtained the same results as shown in the single disk. Therefore, it is still valid to introduce the DM interaction module into the OOMMF code.

3. Results and discussion

To systematically study the effect of the DM interaction on a magnetic vortex, we first simulated the stable vortex magnetic configuration with up polarity and counterclockwise chirality for $D=0$, and then studied how the DM interaction influenced the magnetic configuration by gradually varying the $D$ value. Fig. 1 shows the simulated magnetic configuration with different $D$ values. There is a phase transition from a vortex state to a helical stripe state when the DM strength reaches a threshold $D_{crit}$. This threshold depends on the disk parameters, and the $D_{crit}$ value in this simulation is 1.76 mJ/m². When $|D|>D_{crit}$, the DM energy is strong enough to overcome the dipole energy in the system, so then the interplay between the DM energy and the exchange energy forms a helical spin structure, which is close to the helical stripe phases in 2D thin film [6]. For $|D|<D_{crit}$, the disk still keeps the vortex configuration, but a positive $D$ can widen the core, while a negative $D$ shrinks the core. This phenomenon is consistent with the previous study based on the analytical model [19]. Moreover, if we continue to reduce the negative D value, the vortex core polarity can finally be switched by the DM interaction, as shown in Fig. 1(e). This fact means that there only exists the vortex with typical handedness while the DM interaction is sufficiently strong. In the simulation, the critical value to switch the polarity is $D_{switch}$=-1.1mJ/m$^2$, thus only the single handedness vortex could be observed for $|D_{crit}|>|D|>|D_{switch}|$. If the negative D value is further reduced, the DM interaction will increase the size of the vortex core with the reversed polarity, until the vortex state breaks into the helical strip phase (Fig. 1(f)) for the strong negative $D$ value. In Ref. 19, Butenko et. al mentioned that the radial stable solutions exist only below certain critical strength of the DM constant, which may be related to the threshold from the vortex state to the helical stripe state based on the results in Fig.1.

Our simulation further show that the vortex core size depends on the DM interaction nearly linearly, as shown in Fig. 2(a), where the vortex core size is characterized by the width at half maximum of vortex core through the line profile across the vortex core. The vortex core expands from 22 nm to 38 nm as $D$ increases from 0 to 1.6 mJ/m$^2$, but shrinks continually to 8 nm as $D$ drops to -1.1 mJ/m$^2$, below which the core polarity flips from up to down, forming a left-handed vortex. The simulation with smaller unit cell size of $1\times1\times50 nm^3$ shows similar result, and only small difference of core size can be found that for the core size less than 10nm, as shown in Fig. 2(a).

The DM interaction can not only influence the vortex core size, but also influence the magnetization configuration at the disk edge and inside the disk. Fig. 3(a) shows the line profiles across the vortex core with different DM interaction values. The effect of the DM interaction on the core size could be clearly identified through the line profiles. Generally in a system with D=0, the spins apart from the core lie in the film plane due to the in-plane demagnetization field. With the existence of the DM interaction, we found the magnetization at the disk edge could be titled away from the surface plane. The out-of-plane component $M_z$ at the disk edge increases linearly with the DM interaction, and reversed its sign once D changes from positive to negative, as shown in Fig.3b. It should be noted that the edge $M_z$ component decays rapidly within 30 nm away from the disk edge, thus the tilted edge magnetization induced by the DM interaction can only be observed experimentally by those modern magnetic imaging technologies with high spatial resolution, such as magnetic force microscope [22], or spin polarized scanning tunneling microscope [21]. Moreover, we found that the DM interaction could induce a weak out-of-plane magnetization component even in the disk plane. The inset in Fig.3a shows the amplified magnetization profiles with different D values, which clearly proves the $M_z$ components depend on the sign of the DM interaction.

Fig.3b shows that $M_Z$ at $r = 150 nm$ also changes linearly with the D value, but $M_Z$ in the disk plane has the opposite sign with much smaller amplitude than at the disk edge.

Recently, Rohart and Thiaville showed that, in ultrathin film nanostructure with out-of-plane anisotropy, the interfacial DM interaction can also bend the magnetization at the edges towards to the in-plane direction at the edges [31]. This is different with our results that the in-plane magnetization was titled to the normal direction by the bulk-like DM interaction. It would be interesting to further check the effect of the interfacial DM interaction on the vortex in the micro-size magnetic disk with the in-plane magnetization.

The influence of the DM interaction on the vortex core can be attributed to the DM field at the core. Equ. 4 points out that the DM field direction at the vortex core is determined by its neighboring spin direction (the vortex chirality) and the sign of $D$. Since the vector $\vec{r}_{ij}$ is always in the film plane, the in-plane magnetization of the neighboring spins would induce an effective perpendicular DM field. For a vortex with the counterclockwise chirality, the perpendicular component of the DM field at the core is upward when $D > 0$ (see Fig. 2 (b)) and downward when $D < 0$ (see Fig. 2 (c)), thus the vortex core expands when its polarity is parallel to the DM field, and shrinks when its polarity is opposite to the DM field. The internal DM field at the vortex core also lifts the energy degeneration between the left-handed and right-handed vortices, because vortex states have lower energy if the core polarity is parallel to the DM field. In this way, a coupling effect between the vortex polarity and chirality can be expected.

The tilting of the edge magnetization can also be understood by the DM field at the edge. Equ.4 indicates that the in-plane magnetization of neighboring spins can induce an out-of-plane DM field. The magnetization at the disk edge only contains the neighboring spins inside the disk, thus the DM interaction from the magnetizations of inner neighbors will generate the perpendicular DM field on the edge magnetization, and tilt the magnetization away from the surface plane. But this DM field at the disk edge has the opposite sign as that at the core shown in Fig. 2, so the positive D value will induce a negative $M_Z$ at the edge, and the negative D value will induce a positive $M_Z$. The strength of the DM field should depend on the D value, thus the edge magnetization changes almost linearly with the D value.

The out-of-plane component of the magnetization inside the disk is also attributed to the DM interaction. For the magnetization inside the disk, the DM field generated from the inner neighboring spins has the opposite sign of that from the outer neighboring spins. The DM field induced by the outer spins is slightly larger than that induced by the inner spins, so that the overall perpendicular DM field has the opposite sign as that at the disk edge, and the out-of-plane magnetization in the disk has the opposite dependence on the DM interaction, as shown in Fig.3b.

The weak out-of-plane magnetization in the disk can be understood in an analytical way. The DM field in the disk induced by the in-plane curling magnetization can be calculated quantitatively from Equ. 5. If assuming the magnetization outside the vortex core is lying in the film plane and rotating around the disk center, the estimated out-of-plane DM field is $|\overrightarrow{H_{DM}}| = \frac{2D}{\mu_0 M_s r}$ with $r$ representing the distance to the disk center. This DM field will induce a weak $M_Z$ against the in-plane demagnetization field $\overrightarrow{H_d}$. Usually the demagnetization field in the film can be estimated as $|\overrightarrow{H_d}| = M_s$ which could be much larger than $|\overrightarrow{H_{DM}}|$, and thus the out-of-plane magnetization component can be estimated as

$$M_z \approx \frac{|\overrightarrow{H_{DM}}|}{|\overrightarrow{H_d}|} \approx \frac{2D}{\mu_0 M_s^2 r} \qquad (6)$$

So the $M_z$ induced by the DM interaction should be proportional to $\frac{1}{r}$ with a slope of $\frac{2D}{\mu_0 M_s^2}$. This relation is not valid near the vortex core and the disk edge, where the titling angle is so large that the exchange interaction and the long-range dipolar interaction can't be omitted. As an example, we found that the dipolar interaction from the perpendicular magnetization at the core and at the edge induced a negative field on the neighboring spins, such that the magnetization near the core and the edge always shows a negative dip, as shown in fig.3a. In order to reduce the influence from dipolar interaction from the magnetization at disk edge and disk center, we performed a simulation on a magnetic disk with a large diameter of 1502nm and $D = 0.6 mJ/m^2$. Fig.4 shows the calculated magnetization profile, and $M_z$ at the core and at the edge are very close to those in the disks with the smaller diameters, and $M_z$ at the edge also shows a negative value. The $M_z(r)$ with r in the range between 100 nm and 550 nm can be well fitted with a $\frac{1}{r}$ function, as shown in the inset of Fig.4. The fitted coefficient is $1.39 nm^{-1}$, which is very close to the theoretical value of $\frac{2D}{\mu_0 M_s^2} = 1.49 nm^{-1}$. So Equ. 6 can effectively describe the out-of-plane magnetization component in the disk with a large enough diameter.

Fig. 2 shows that the in-plane curling magnetization could induce a perpendicular DM field on the vortex core, so that the DM field would be expected to induce a bias field while the core polarity is switched by the out-of-plane magnetic field. In order to better illustrate the reversing process of the vortex core, we only present the hysteresis loop of the small selected area around the vortex core, as shown by the green rectangle area ($62 \times 62 nm^2$) in Fig. 5(a). Fig. 5(b) shows the typical hysteresis loop of the selected area on the disk with $D = 0.2 mJ/m^2$. The applied magnetic field is strong enough to saturate the magnetization along the normal direction, as shown by the insets in Fig. 5(b). The magnetic configurations at the remanence show clear vortex states, however for the field sweeping downward, the vortex always has up polarity and counterclockwise chirality, and for the field sweeping upward from a negative saturation field, the vortex has down polarity and clockwise chirality. It is clear that the vortex core polarization can be determined by the applied field direction during the vortex creation process, and thus the simulation results indicate that the chiral direction always follows the core polarization in the vortex-formation process. So the DM interaction couples polarity and circularity of the vortex in magnetic microdisk, and the right-handed vortex state is the energy favorable state for positive $D$ value. We also performed the simulation without the DM interaction, and did not observe the coupling effect between the polarity and the circularity of the vortex. In the magnet vortex induced by the DM interaction in the nanodisk with the radius less than 30nm, Kwon et. al [20] also showed that the polarity and circularity of the skyrmion structure are coupled together. through the field dependent simulation.

As shown in Fig.2, the DM interaction can provide a DM field on the vortex core. This DM interaction not only modifies the vortex core size, but also is the origin to couple the polarity and the circularity of the vortex. We further understood the handedness preference of the vortices induced by the DM interaction more clearly through the minor hysteresis loop, in which the magnetic field is only strong enough to switch the vortex core polarization without changing the vortex circularity. In this case, the applied field should be smaller than the saturation field, i.e. less than 860 mT in the simulation. Fig. 5(c) show the typical minor loops of the selected area with different DM interaction constants. The insets of Fig.5(c) show that only the core polarities can be switched without changing the counterclockwise chirality during the field sweeping process. For the conventional vortex without DM interaction, the obtained loop is symmetrical, and the vortex core switching field $H_c^-$ for the polarization from up to down and the switching field $H_c^+$ for the polarization from down to up have the same magnitude of 650mT. However, for the vortex with the DM interaction, a clear offset of the switching fields can be observed. When $D = 0.2 mJ/m^2$, $H_c^+$ is 550mT, and $H_c^-$ is -750 mT, so that this simulation confirmed that there is a positive bias field of 100mT on the vortex core induced by the DM interaction, which is consistent with the physical picture in fig. 2. It requires an extra field to overcome the bias field for the core polarization switching from up to down. We found that the bias field can reverse its sign after the DM interaction becomes negative, and is proportional to the DM interaction. Thus the observed bias field is a clear evidence to prove the existence of the effective DM field induced by the DM interaction. However, we can only observe the biased vortex core loops for the DM interaction weaker than $0.4 mJ/m^2$, because for stronger DM interaction, the core cannot be switched when the applied field is less than the saturation field. Here it should be noted that the vortex core switching field could be increased by choosing smaller unit cell size in the simulation [32], but the bias field induced by the DM interaction has little dependence on the unit cell size.

In magnetic disks, it is difficult to control the vortex circularity, and usually the asymmetry disks, such as edge-cut disks [33,34], were applied to control the vortex chirality. Our simulation clearly demonstrated that the DM interaction can couple the polarity and chirality together, so that the vortex chirality in a circle disk can be controlled with its polarity by switching the external field. Although the ordinary material used in the study on magnetic vortex, such as permalloy, is unlikely to contain large DM interaction, the DM strength in the B20 materials, such as $Fe_{0.5}Co_{0.5}Si$, can be up to 0.48 mJ/m$^2$ [6, 20], and the discovered DM-interaction-induced effect is feasible to be realized experimentally in the magnetic nanodisk made by the B20 materials.

4. Summary

We have studied the effect of the DM interaction on magnetic vortices by micro-magnetic simulation based on the LLG equation. The DM interaction in magnetic mcrodisks can influence the size of the vortex core, and destabilize one vortex handedness at intermediate DMI strength, and destabilize all vortex states into the helical stripe phase for strong DMI strength. The DM interaction can also induce an out-of-plane magnetization component at the edge and an opposite component at the disk plane. We found that the effective DM field could induce a bias field on the vortex core while switching the core polarization with an out-of-plane magnetic field, and further induce the coupling between the vortex circulation and polarity. Our calculations indicate that the DM interaction can be a new and efficient way to control the vortex in magnetic microdisks.

We acknowledge helpful discussions with Prof. Shufeng Zhang and Prof. Haifeng Ding. This work was supported by MOST (No. 2011CB921801, No. 2009CB929203), by NSFC of China (No. 10925416 and No. 11274074), by WHMFC (No. WHMFCKF2011008), and by the National Research Foundation of Korea Grant funded by the Korean Government (2012R1A1A2007524).

Figures:

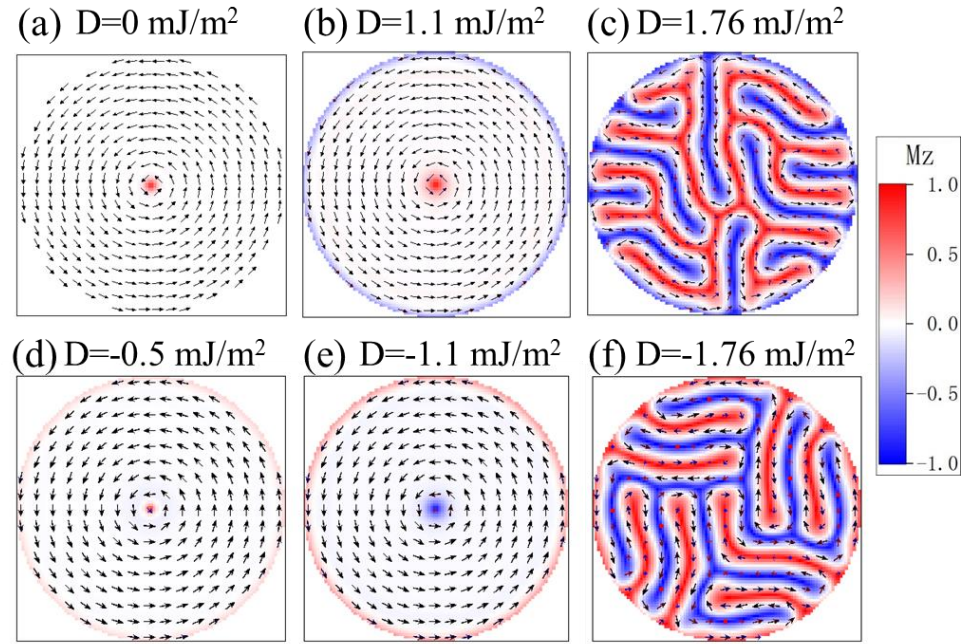

Fig. 1. Magnetization configurations in the vortices with different $D$ values. The color represents the out-of-plane magnetization direction as indicated by the right color bar; the arrows denote the in-plane magnetization direction.

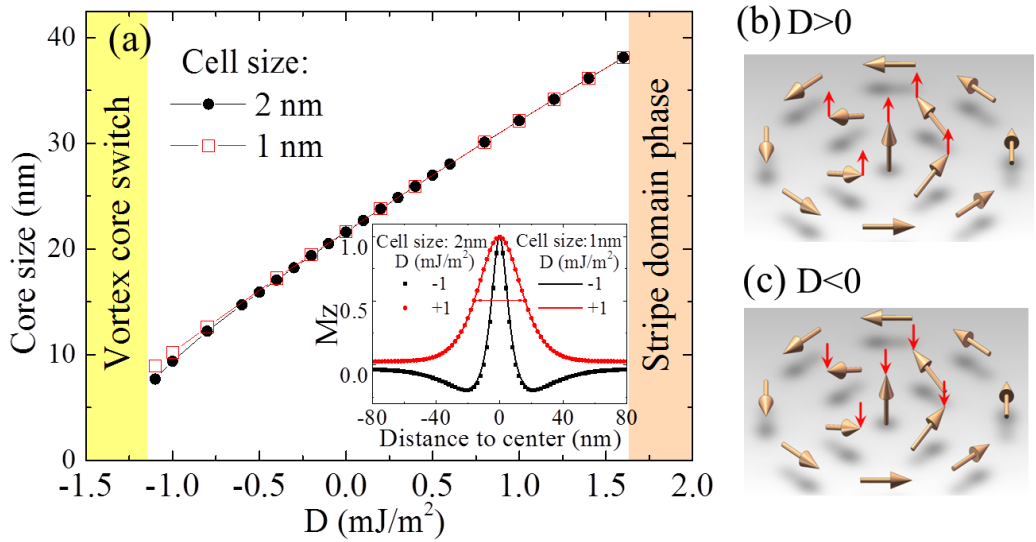

Fig. 2. (a) The size of the vortex core as a function of the $D$ value. The inset shows the typical $M_z$ line profile across the vortex core with two representative D values, $r$ denotes the distance to the center. The simulation was performed with the unit cell size of 2nm and 1nm respectively. The size of the vortex core is defined as the peak width at $M_z = 0.5$ indicated by the red arrow. (b) and (c) show schematic drawings of the DM field at the vortex core for (b) $D>0$ and (c) $D<0$. Yellow arrows represent the spin structures around the vortex core, and the red arrows denote the out-of-plane component direction of the DM field.

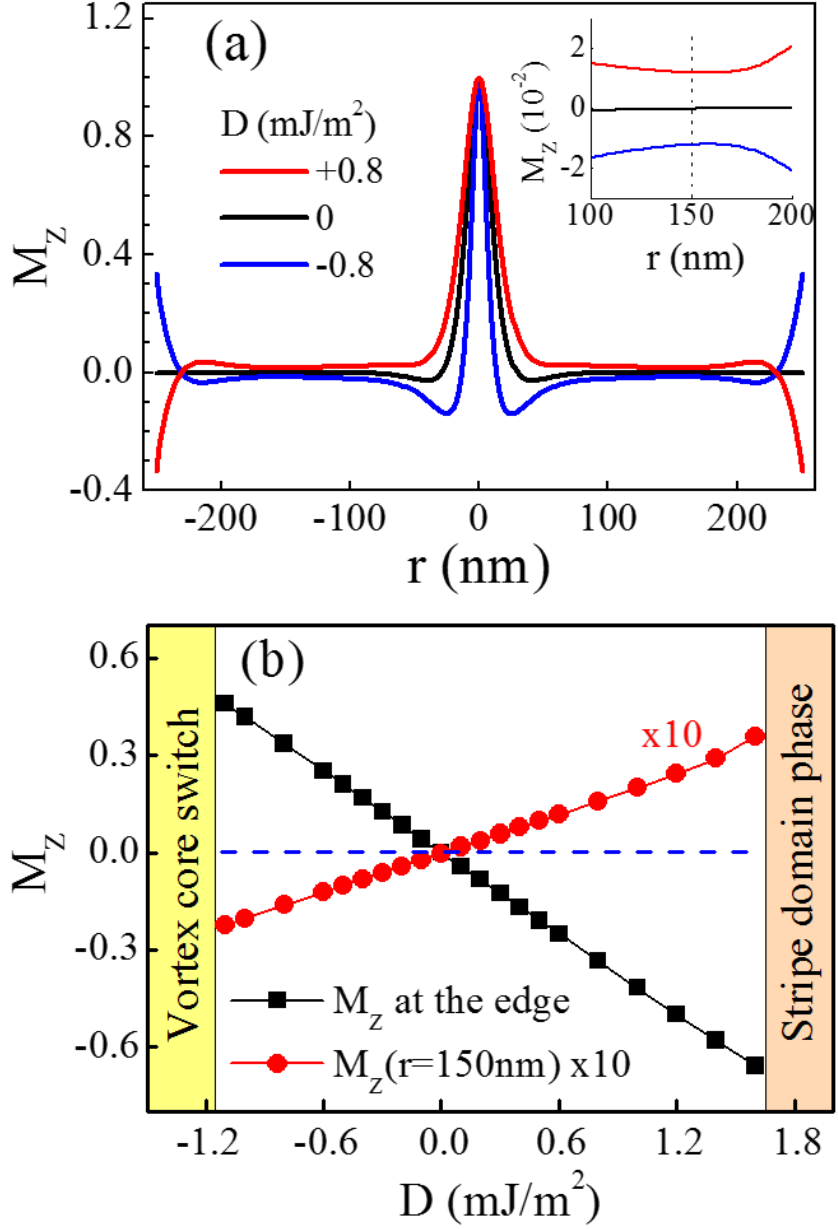

Fig. 3. (a) $M_z$ line profile across the vortex core with different D values, $r$ denotes the distance to the center. The insert shows the magnified $M_z$ profile. (b) The $M_Z$ at the disk edge and at r=150 nm as a function of the DM interaction constants.

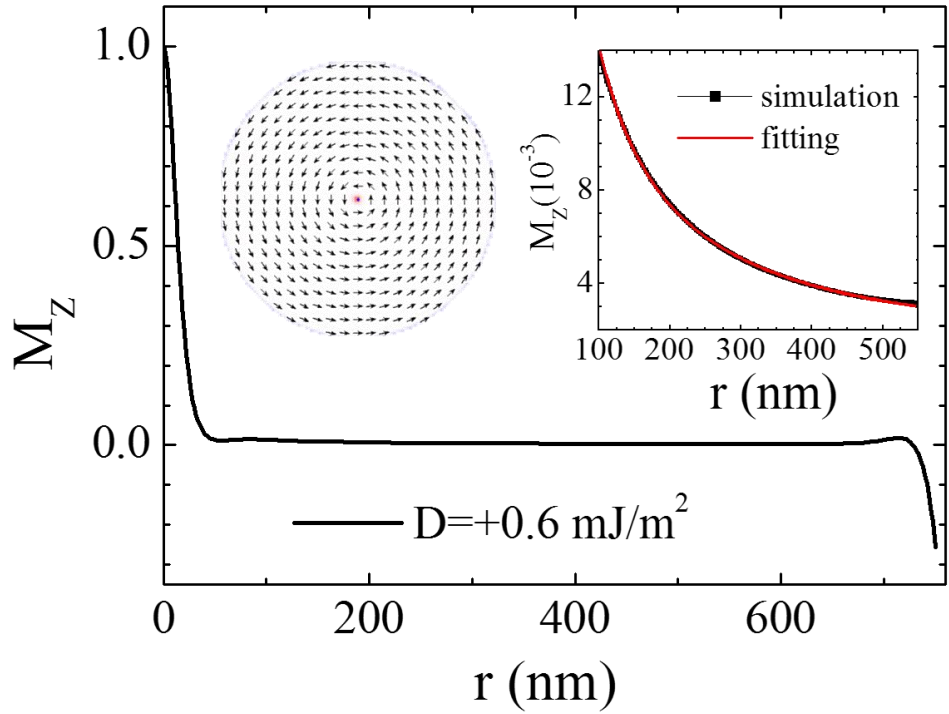

Fig. 4. $M_z$ line profile of a disk with a diameter of 1502nm and a D value of 0.6 mJ/m², $r$ denotes the distance to the center. The left insert shows the magnetic configuration of the disk, and the right insert shows the magnified line profile and the fitting curve from the $1/r$ function.

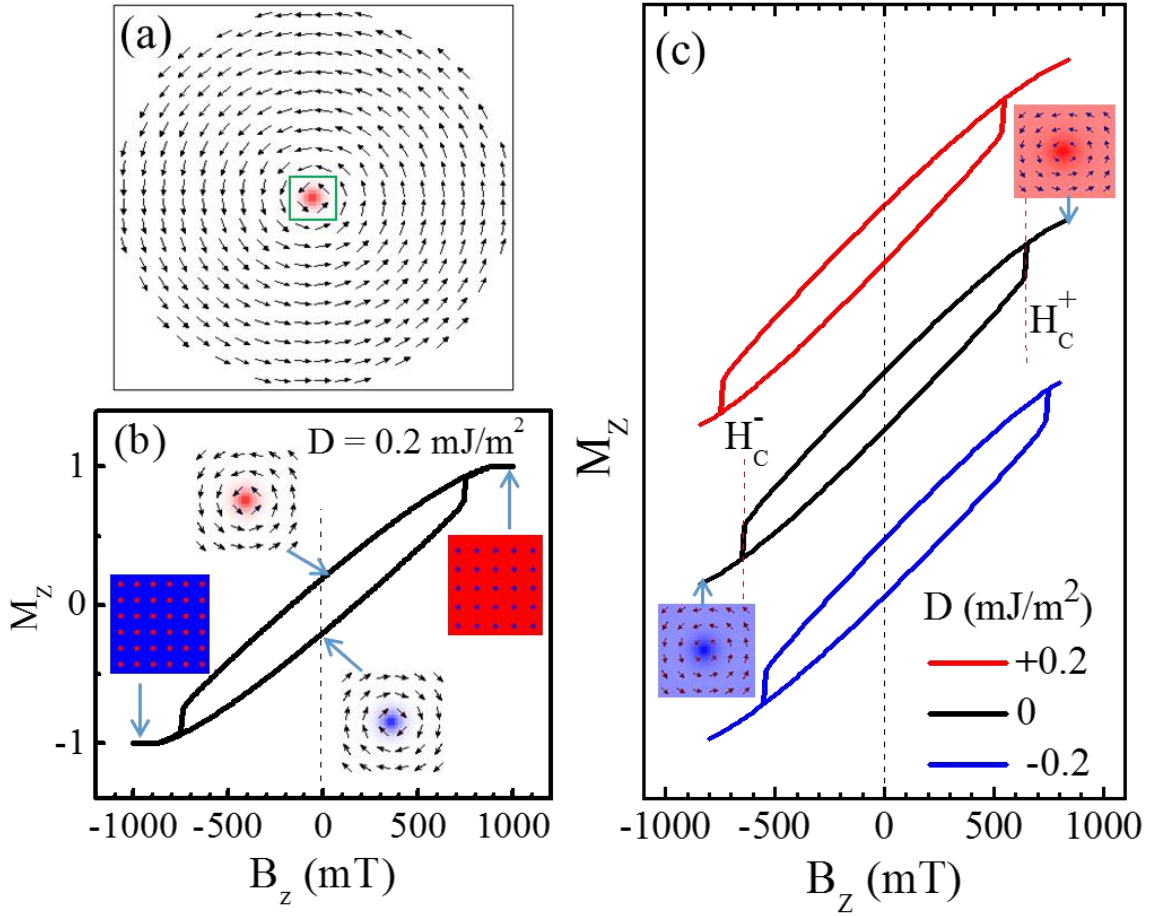

Fig. 5. (a) The selected green rectangular area ($62 \times 62 nm^2$) around the vortex core. (b) The magnetic hysteresis loop of the selected area in the vortex with $D = 0.2 mJ/m^2$. The insets show the magnetization configurations inside the selected area at saturation states and remanence states. (c) The minor loops of the selected area with different $D$ values. The insets show the magnetization configurations at 800mT and -800mT with the same chirality.

*Correspondence to: wuyizheng@fudan.edu.cn